\documentclass{aa}
\usepackage{graphicx}

\usepackage[german,english]{babel}
\begin{document}

\thesaurus{08(02.01.2;08.02.1;08.09.2\,AM\,Her;08.14.2)}

\title{The response of a dwarf nova disc to real mass transfer variations}
\author{M.R. Schreiber, B.T. G\"ansicke \and F.V. Hessman}
\institute{Universit\"ats-Sternwarte, Geismarlandstr.11, D-37083 G\"ottingen, Germany}

\authorrunning{Schreiber et al.}

\maketitle

\begin{abstract}
We present simulations of dwarf nova outbursts taking into account
realistic variations of the mass loss rate from the secondary.
The mass transfer variation has been derived from 20 years of visual
monitoring and from X-ray observations covering various
accretion states of the discless cataclysmic variable AM\,Herculis.
We find that the outburst behaviour of a fictitious dwarf nova with the same
system parameters as AM\,Her is
strongly influenced by these variations of the mass loss
rate. Depending on the mass loss rate, the disc produces either 
long outbursts, a cycle of one long outburst followed by two short
outbursts, or only short outbursts.
The course of the transfer rate dominates the shape of the
outbursts because the mass
accreted during an outburst cycle roughly equals the mass
transferred from the secondary over the outburst interval. Only
for less than 10\% of
the simulated time, when the mass
transfer rate is nearly constant, the disc is in a
quasi-stationary state during which it periodically repeats the same cycle
of outbursts.
Consequently, assuming that the secondary stars in non-magnetic CV's
do not differ from those in magnetic ones, our simulation indicates that
probably all dwarf novae
are rarely in a stationary state and are constantly adjusting to the
prevailing value of the mass transfer rate from the secondary. 


\end{abstract}

\section{Introduction}
Cataclysmic variable systems (CV's), in which a white dwarf accretes
material
from a Roche lobe filling secondary star (see Warner 1995 for an
enyclopaedic review) can be broadly divided into three
subclasses: in non-magnetic systems, the white dwarf accretes via an
accretion disc; in the magnetic polars, the infalling matter
couples directly onto the strong magnetic field of the white dwarf before
it can build a disc and is
funnelled to accretion region(s) near the magnetic pole(s) of the white
dwarf; finally, in CV's containing a weekly magnetic white dwarf
(intermediate polars), a partial disc may exist, with the mass flowing
from the inner edge of the disrupted disc through magnetically funnelled
accretion curtains onto the white dwarf.

Dwarf novae are non-magnetic cataclysmic variable stars which show
characteristic eruptions with photometric amplitudes in the range of 2--8
magnitudes {\footnote{Also intermediate polars are known to show outbursts
in their partial discs; e.g. EX Hya, GK Per}}. These eruptions typically
last a few days to several weeks and recur quasi-periodically
on timescales of weeks up to many years. They are generally
thought to result from thermal instabilities associated with hydrogen
ionization in the accretion disc (see Cannizzo 1993a for a review and
Ludwig et al. 1994 for a detailed parameter study).

An important boundary condition
of the disc-limit-cycle model is the mass transfer rate from the
secondary to the accretion disc. In most studies of dwarf nova
outbursts this mass transfer rate is kept constant (Cannizzo 1993b;
Ludwig \& Meyer 1997; Hameury et al. 1998, 1999). Duschl and
Livio (1989) were the first to examine combined mass transfer and
disc outbursts, though within the framework of the mass transfer
instability model where individual and short-lived mass transfer events
are capable of producing single outbursts. Smak (1991, 1999) discussed
mass transfer variations in the context of the superoutburst phenomenon in
dwarf novae of the SU\,UMa type and dwarf nova outbursts with
enhanced mass transfer during outburst.

King \& Cannizzo (1998) and Leach et al. (1999) tested how the
accretion disc in a dwarf nova system behaves if the mass transfer
from the secondary varies abruptly between different
levels.  They found that these mass transfer variations produce only
subtle effects on normal dwarf novae, including variations in the outburst
shape, and that the dwarf novae keep on having outbursts even if
the transfer rate is reduced close to zero.

In spite of these efforts we are left with the question of what the
real variations of the mass tranfer rates from the secondaries in
dwarf nova systems are and how they can influence the
outburst behaviours of the accretion discs. Fortunately,
nature provides an answer to the first question in the form of
discless cataclysmic variables, the polars or AM\,Her systems.
In these systems,
the mass transfer rate can be estimated directly from the
observations because there is no accretion disc acting as a
mass buffer.
Thus, we can use the long-term light curve of an AM\,Her system
as a measure for the mass transfer variations in a fictitious but realistic
dwarf nova system with the same system parameters.
In the present paper, we present the results of such a numerical experiment.

In the following section, we briefly review the equations for the
viscous and thermal evolution of the accretion disc, discuss the
numerical methods used in our code, and compare the results of two
standard models with other fine mesh computations.  Thereafter we
derive the mass transfer rate in AM\,Her as a function of time,
$\dot M_\mathrm{tr}(t)$. Finally, we apply this mass transfer
rate to a fictitious dwarf nova and discuss
the effects that can be observed in the outburst behaviour.

\section{Finite Element Methods in the context of disc evolution}

The classical equation describing the viscous evolution of the surface
density $\Sigma$ in a geometrically thin, axisymmetric accretion disc is
obtained by combining the vertically averaged Navier-Stokes and
mass-conservation equations:
\begin{eqnarray}			\label{sigmaEqn}
\frac{\partial \Sigma}{\partial
t} &=& \frac{3}{R}\frac{\partial}{\partial
R}\Big(R{^\frac{1}{2}}\frac{\partial}{\partial R}
	(R^{\frac{1}{2}}\nu\Sigma)\Big)+\frac{\dot M_\mathrm{tr}(t,R)}{2\pi R}
\end{eqnarray}
where $\nu=(2/3)\alpha R_{\mathrm{g}}T/\mu \Omega$ is the kinematic
viscosity, $\alpha$ denotes the viscosity parameter, $R_{\mathrm{g}}$ the gas constant and $\mu$ the mean molecular weight. The first term on the right hand side is the
classical disc diffusion term and the second term describes the
external mass-deposition. For the purposes of this study, we follow
Cannizzo (1993b) and simply
add the mass lost from the
secondary $\dot M_\mathrm{tr}$(t) in a Gaussian distribution at a fixed radius near the outer edge.

Similarly, one can derive an energy equation (in the central temperature $T$:
\begin{equation}
\frac{\partial T}{\partial t}=\frac{2(H-C+J)}{c_{\mathrm{p}} \Sigma}-\frac{R_{\mathrm{g}}T}{\mu c_{\mathrm{p}}}
	\frac{1}{R} \frac{\partial (R v_{\mathrm{R}})}{\partial R}
	-v_{\mathrm{R}}\frac{\partial T}{\partial R},
\end{equation}
where $c_{\mathrm{p}}$ denotes the specific heat, 
\[
v_{\mathrm{R}}=\frac{-3 \nu}{R} \frac{\partial \log(\nu \Sigma
r^{\frac{1}{2}})}{\partial \log R}
\]
is the local radial flow velocity, $H=\frac{9}{8}\nu \Omega^2 \Sigma$
represents viscous heating, $C=\sigma T_\mathrm{eff}^4$
is the radiative cooling and 
\[J=\frac{3}{2}c_{\mathrm{p}} \nu
\frac{\Sigma}{R}\frac{\partial}{\partial R}(R\frac{\partial T}{\partial R})\]
the
radial energy flux carried by viscous processes (see Smak 1984; Mineshige \&
Osaki 1983; Mineshige 1986; Ichikawa \& Osaki 1992 and Cannizzo 1993b, for
discussions). 

We solve Eqs. (1) and (2) using a combined Finite-Element /
Finite-Difference algorithm (FE for the spatial part and FD for the
time-evolution).  Apart from our own work (Schreiber \& Hessman 1998)
the method of Finite Elements has not been used in this
context. As this method proved to be
extremely robust, it warrants a somewhat more detailed description.

The idea of FE is to divide the region of interest (the disc radii between
$R_{\mathrm{in}}$ and $R_{\mathrm{out}}$) into $n-1$ elements and to expand the function $u(x)$ which
is supposed to solve the differential equation with suitable functions
$u(x)=\sum_{i=1}^{n} a_i \varphi_i(x)$ for every element.
In order to get
a continuous solution over all elements, the functions ($\varphi$) of
every element have to be transformed to the so-called local basis
functions $N_{i}$ and the coefficients to the so-called nodes
$(c_{i})$ before collected together (Gruber \& Rappaz 1985, Schwarz 1991).
To solve the differential equation, the function
\begin{equation}
u(x)=\sum_{k=1}^n c_i N_i(x)
\end{equation}
has to meet the requirement formulated by
Garlerkin: the integral of the residuum (which one gets by
inserting Eq.\,(3) into the differential
equation) weighted with the functions $N_j(x)(j=1,...,n)$ has to vanish.
This requirement, the interchange of integration and summation and
partial integration lead to matrix-equations of the form
\begin{equation}
B\dot{c}+Ac = D,
\end{equation}
with $A=(a_{ij}), B=b_{ij}$ and $D=d_i$ ($i=1,..,n; j=1,..,n$ for $n$ nodes).

To solve the differential Eqs.\,(1) and (2), we have to fill $A$, $B$, $D$
in the sense mentioned above and calculate $c$ from Eq.\,(4).   
After transforming to the variables $X=2R^{\frac{1}{2}}$ and $S=X\Sigma$ we derive for the surface density from Eq.\,(1):
\begin{eqnarray}
\nonumber
a_{ij} &=& \sum_{k=1}^{n} \nu_k \Big( \int \frac{12}{X^2}\frac{\partial N_i}{\partial
X}N_k \frac{\partial N_j}{\partial X}dX\\ \nonumber
 & & \hspace{0.8cm}+\int \frac{12}{X^2}N_i \frac{\partial N_k}{\partial
X}\frac{\partial N_j}{\partial X}dX\Big)\\ \nonumber
b_{ij} &=& \int N_i N_j dX\\ \nonumber
d_i &=& \sum_{k=1}^{n} \int \frac{2\dot M_{\mathrm{tr},k}N_k}{\pi X^2}N_jdX.
\end{eqnarray}  
Similarly it is easy to obtain for the central temperature Eq.\,(2): 
\begin{eqnarray}
\nonumber
a_{ij} &=& \sum_{k=1}^{n} \Big( \int p^{(1)}_k\frac{\partial N_i}{\partial
R}N_k \frac{\partial N_j}{\partial R}dR\\ \nonumber
& & \hspace{0.2cm}+\int p^{(2)}_k N_k \frac{\partial N_i}{\partial R} N_j dR+\int p^{(3)}_k N_i N_k N_j dR \Big) \\ \nonumber
b_{ij} &=& \int N_i N_j dR\\ \nonumber
d_j &=& \sum_{k=1}^{n} \int p^{(4)}_k N_k N_jdR.
\end{eqnarray}  
The coefficients $p^{(i)}_k$ are given by
\begin{eqnarray}
\nonumber
p^{(1)}_k &=& 3c_{\mathrm{p}} \nu \Sigma\\ \nonumber
p^{(2)}_k &=& \frac{3c_{\mathrm{p}} \nu \Sigma}{R}-v_{\mathrm{R}}\\ \nonumber
p^{(3)}_k &=& \frac{R_{\mathrm{g}}}{\mu c_{\mathrm{p}} R}\frac{\partial Rv_{\mathrm{R}}}{\partial R}\\ \nonumber
p^{(4)}_k &=& \frac{9/4 \nu \Sigma \Omega^2-\sigma T_\mathrm{eff}}{c_{\mathrm{p}} \Sigma}.
\end{eqnarray}
The index $k$ refers to the value of the quantities at node number
$k$, which is equivalent to the radius $R_k$.
Notice that we only used linear basis functions in this paper.   

As mentioned above, Eq.\,(4) has to be solved to get
$c(t+\Delta t)$ from $c(t)$. Using simple
finite differences leads to
\begin{eqnarray}c(t+\Delta t) &=& c(t)\frac{1}{2}\Delta 
t(B^{-1}Ac(t) \nonumber\\
& & +B^{-1}Ac(t+\Delta t)(+B^{-1}D)).\end{eqnarray}

\begin{figure}
\includegraphics[angle=270,width=9cm]{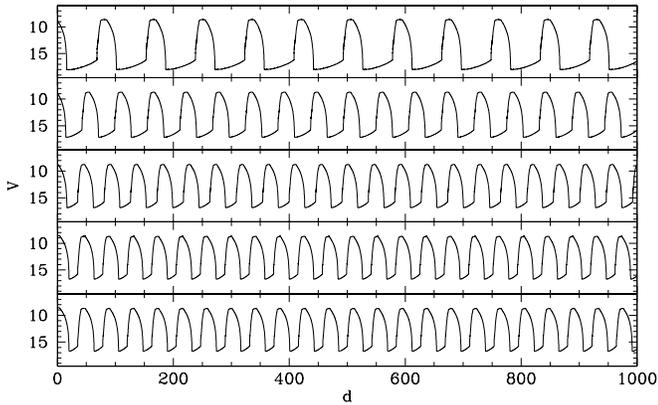}
\caption[]{\label{cannizzo}Comparison with Cannizzo (1993b, his Fig.6d). From top to bottom we used 40, 60, 80, 100 and 200 nodes. Convergence is obtained for
100 nodes.}
\end{figure}

\begin{figure}
\includegraphics[angle=270,width=9cm]{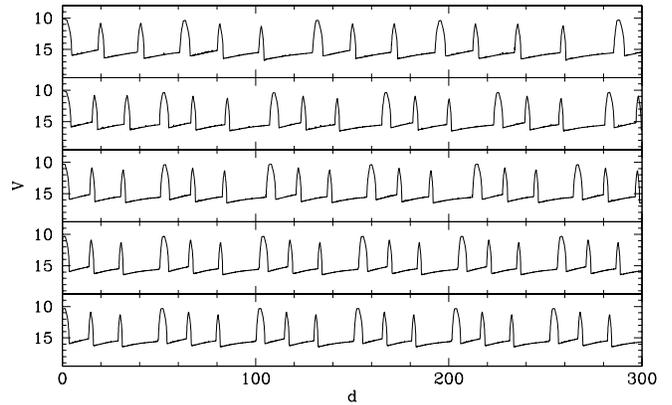}
\caption[]{\label{ludwig} Comparison with Ludwig \& Meyer (1997). From
top to bottom we again used 40, 60, 80 ,100 and 200 nodes. Our code acurately
reproduces the sequence of one long outburst followed by two
short outbursts.}
\end{figure}

In order to test our code, we carried out two sets of calculations
using the binary parameters and cooling functions from Cannizzo
(1993b)
\begin{eqnarray*}
& & M_1 = 1M_{\odot},\, \alpha_{\mathrm{h}}=0.1, \,\alpha_{\mathrm{c}}=0.02, \\
& & R_\mathrm{in}=5.0\times 10^8 \mathrm{cm},\, 
    R_\mathrm{out}=4.0\times 10^{10}\mathrm{cm}, \\
& & \dot M_\mathrm{tr}=1.5\times 10^{9} M_{\odot}/\mathrm{yr}
\end{eqnarray*}
and Ludwig \& Meyer (1997)
\begin{eqnarray*}
& & M_1 = 0.63 M_{\odot},\, \alpha_h=0.2, \alpha_c=0.04, \\
& & R_\mathrm{in} = 8.4\times 10^8 \mathrm{cm},\, R_\mathrm{out}=1.7\times 10^{10} \mathrm{cm}, \\
& & \dot M_\mathrm{tr} = 5\times 10^{15} \mathrm{g/s}.
\end{eqnarray*}
The resulting light curves are shown in 
Figs.\,\ref{cannizzo} and \ref{ludwig}. 
Our code reproduces the sequence of only relatively long outbursts
found by Cannizzo (1993b) for the parameter of SS\,Cygni as well as the
sequence of one long outburst followed by two short outbursts found by
Ludwig \& Meyer (1997) to describe VW Hydri. The short outbursts arise
when there is not enough mass stored in the disc and therefore the
heating wave gets reflected before it has reached the outer edge of
the disc.

We find that at least $100$ nodes are necessary for long-term convergence.
The outburst and quiescence
duration decreases with an increasing number of nodes because - with
finer zoning - the length of time spent on the viscous plateau becomes
shorter (Cannizzo 1993b). This effect is smaller in Fig.\,\ref{ludwig}
because the disc is smaller in this system.

We conclude that our FE-code produces results which are in excellent agreement
with those of other fine-mesh computations.

\section{The mass loss rate of the secondary star in AM\,Herculis}
\begin{figure*}
\includegraphics[angle=270,width=18cm]{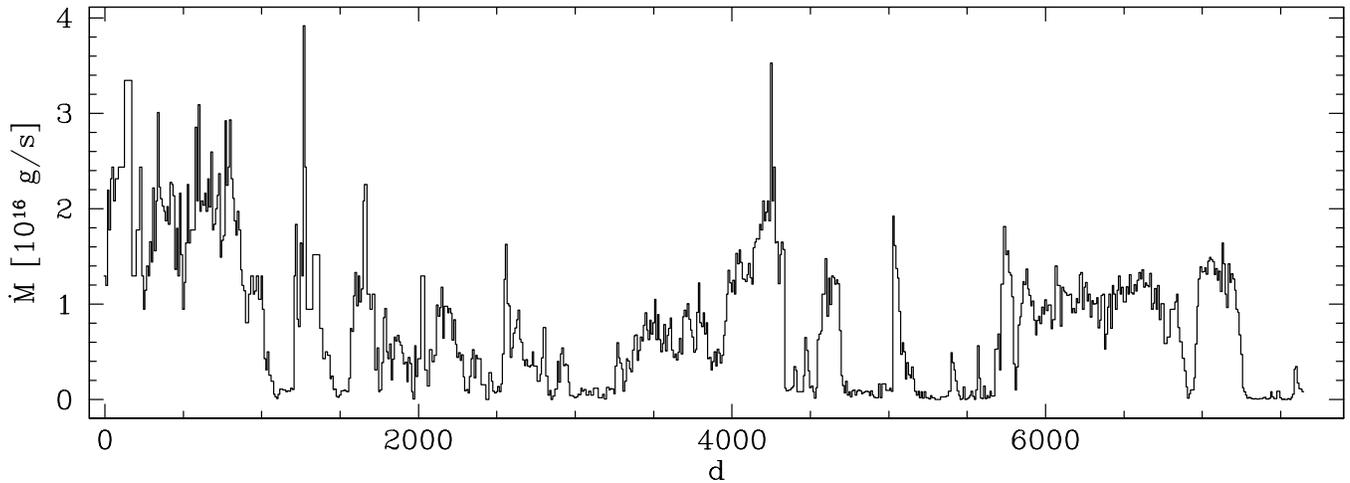}
\caption[]{\label{f-mdot} The mass transfer rate in AM\,Her as a
function of time.}
\end{figure*}

The strong magnetic field of the white dwarf
primary in polars prevents the formation of an accretion disc. Without
an accretion disc acting as a buffer for the transferred mass, the
mass loss rate from the secondary equals the mass accretion rate on
the white dwarf {\it at every moment}, $\dot M_\mathrm{acc}=\dot
M_\mathrm{tr}$ (the free-fall time is $\la1$\,h).
As an observational consequence, any variation in rate at which
the secondary star loses mass through the $L_1$ point will result in a
quasi-immediate change of the observed accretion luminosity. The
brightest polar, AM\,Her, has been intensely monitored at optical wavelengths
by observers of the AAVSO for more than 20 years and shows an
irregular long-term variability, switching back and forth between
high- and low states of accretion on timescales of days to months.  

The problem of deriving the mass loss history of the secondary star is
then equivalent to that of  determining the accretion luminosity
$L_\mathrm{acc}$ as a function of time. As the bulk of the accretion
luminosity is emitted in the X-ray regime, a bolometric correction
relating the densly monitored optical magnitude to the total
luminosity has to be derived. This approach has been followed in
detail by Hessman et al. (A\&A, submitted) using X-ray observations
obtained at multiple epochs. We summarize only briefly the
results here. The accretion luminosity is computed from the observed
accretion-induced flux $F_\mathrm{acc}$ as
\begin{equation}
L_\mathrm{acc}(t) = 4\pi d^2 F_\mathrm{acc}(t)
\end{equation}
with $d=90pc$ (G\"ansicke et al. 1995). We further use
$F_\mathrm{acc} \approx F_\mathrm{SX} + 3\times F_\mathrm{HX}$ with
$F_\mathrm{SX}$ and $F_\mathrm{HX}$ the observed soft and hard X-ray
fluxes respectively. The factor three accounts for the additional
cyclotron
radiation emitted from the accretion column and for the thermal
reprocession of bremsstrahlung and cyclotron radiation intercepted by
the white dwarf and emitted in the ultraviolet (G\"ansicke et
al. 1995). The mass loss (\,=\,transfer) rate is then
\begin{equation}
\label{mtr}
\dot{M}_\mathrm{tr}(t) = 
\frac{L_\mathrm{acc}(t)R_\mathrm{wd}}{G M_\mathrm{wd}}
\end{equation}
where $G$ is the gravitational constant and $R_\mathrm{wd}$ and
$M_\mathrm{wd}$ are the white dwarf radius and mass, respectively.  As
the actual properties of the white dwarf in AM\,Her are still the subject
of controversial discussions (G\"ansicke et al. 1998, Cropper et
al. 1999), we use the parameters of an average white dwarf,
 $M_\mathrm{wd}=0.6\,M_{\odot}$ and
$R_\mathrm{wd}=8.4\times10^8\,\mathrm{cm}$. $\dot{M}_\mathrm{tr}(t)$ is shown
in Fig.\,\ref{f-mdot}. The average value of the mass transfer rate in
AM\,Her is $\dot{M}_{\mathrm{av}}=7.88\times10^{15}\mathrm{g\,s^{-1}}=
1.24\times10^{-10}M_{\odot}\mathrm{yr}^{-1}$.  The derived mass transfer
rates of AM Her are in general agreement with results published in the
literature. For example, Beuermann \& Burwitz (1995) found transfer
rates between $0.8$ and $2.0\,\times\,10^{-10}\,\dot{M_{\odot}}\mathrm{yr}^{-1}$
and Greeley et al. (1999) estimated a mass transfer rate
of $2\,\times\,10^{16}\,\mathrm{g\,s^{-1}}$ for the
high state of AM\,Her from far ultraviolet spectra.

\section{Results}
\subsection{The fictitious dwarf nova}
We devised a fictitious dwarf nova with a non-magnetic primary of mass
$M_\mathrm{wd}=0.6\,M_{\odot}$ and an orbital period of $P=3.08\,\mathrm{hr}$,
i.e. a non-magnetic twin of AM\,Her.  For these binary parameters, we
obtain $R_\mathrm{in}\simeq R_\mathrm{wd}=8.4\times10^8$\,cm for the
inner disc radius and $R_\mathrm{out}=2.2\times10^{10}$\,cm for the
outer edge of the disc. We then used our FE code to follow the
structure of the accretion disc in our fictitious dwarf nova for
7000\,d, applying the variable mass transfer rate $\dot{M}_\mathrm{tr}(t)$
derived above and standard viscosity parameters
$\alpha_{\mathrm{h}}=0.2$ and $\alpha_{\mathrm{c}}=0.04$.

We show $500$ day-long samples of our calculations in
Figs.\,4--8.
In each figure, the top panel show the mass transfer rate as a function of time (solid
line) and the average mass transfer rate
$\log(\dot{M}_{\mathrm{av}}[\mathrm{g}\,\mathrm{s}^{-1}])=15.90$
(dotted line). The panel below displays the disc mass
$M_{\mathrm{disc}}$ normalized with the averaged disc mass
$\overline{M}_{\mathrm{disc}}=1.64\,\times\,10^{23}\mathrm{g}$. The two lower
panels
display the
light curves
calculated with the varying mass transfer rate and the light curves calculated
with the constant average mass
transfer rate respectively. For the constant average mass transfer rate the
disc goes through a $\sim\,60$\,day-long cycle including one long
outburst followed by two short outbursts.  The long outbursts are
those in which the entire disc is transformed into the hot state while
the short outbursts arise when the outward moving heating wave is
reflected as a cooling wave before it has reached the outer edge of
the disc.

\begin{figure}
\includegraphics[angle=270,width=9cm]{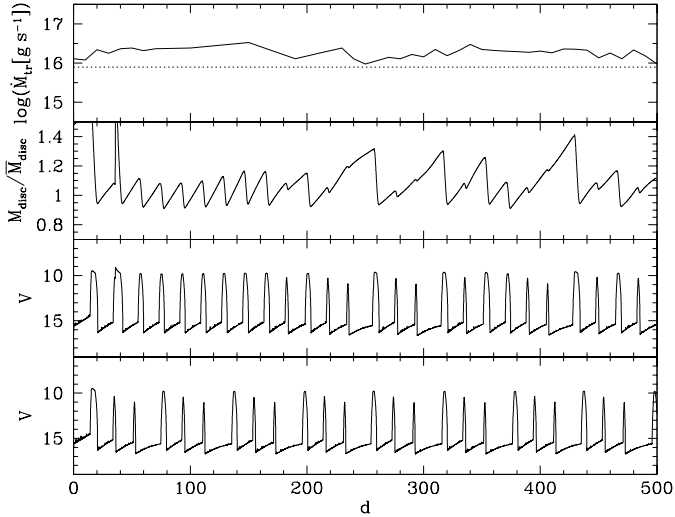}
\caption[]{\label{exam-1}
From top to bottom as a function of time: the mass transfer rate
(top, solid line), the averaged mass transfer rate
(top, dotted line), the normalized disc mass 
and the light curves produced by the fictitious dwarf nova with the
variable transfer rate and the averaged transfer rate adopted from
AM\,Her. }
\end{figure}

Our numerical experiment clearly demonstrates that the outburst
light curve of the fictitious system is strongly affected by the
variations of the mass transfer rate. Even in the case of relatively
small fluctuations effects on the
outburst behaviour
(\mbox{$16.0\,<\,\log(\dot{M}_{\mathrm{tr}}[\mathrm{g\,s^{-1}}])\,<\,16.4$})
are clearly present in the light curve. In Fig.\,\ref{exam-1},
the mass transfer rate is always high during the 500\,days but the
disc switches between an accretion state with only long outbursts and
states where one or two short outbursts follow a long outburst. When
the tranfer rate decreases somewhat ($\sim$\,day\,200), 
the disc does not save enough mass to create consecutive long
outbursts until the mass transfer rate increases again ($\sim$\,day\,350).

\begin{figure}
\includegraphics[angle=270,width=9cm]{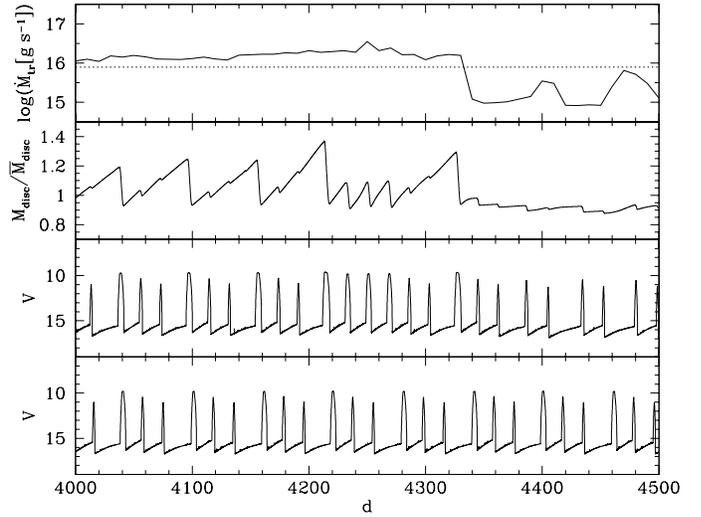}
\caption[]{\label{exam-2} The same as Fig.\,\ref{exam-1} but another
snapshot of the simulation. The sharp decline of the mass transfer
rate is immediately reproduced by the accretion disc.}
\end{figure}

In addition to this effect, our experiment shows that a sharp decrease
in the mass transfer rate instantaneously changes the outburst
behaviour of the accretion disc. In Fig.\,\ref{exam-2} the disc first
behaves as in Fig.\,\ref{exam-1} but when the transfer rate drops
sharply (day\,4340), the long outbursts immediately vanish and the
duration of the quiescent phase increases somewhat.

\begin{figure}
\includegraphics[angle=270,width=9cm]{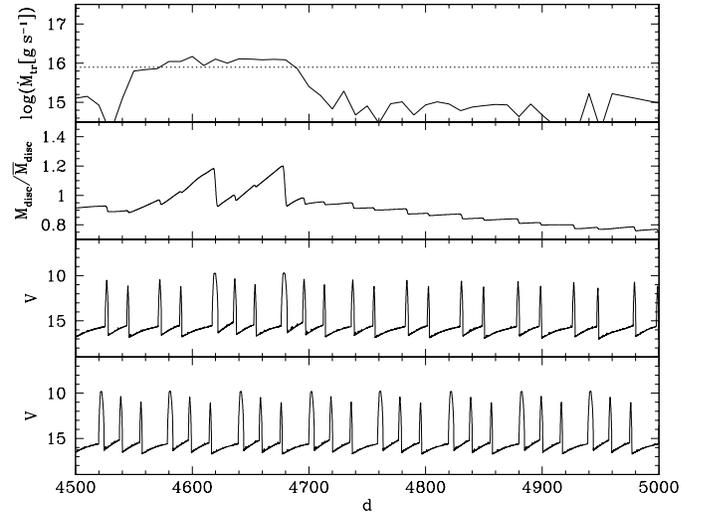}
\caption[]{\label{exam-3} The same as Fig.\,\ref{exam-1} but another
snapshot of the simulation. Even the long period of low transfer rates
does not stop the outburst activity of the disc.}
\end{figure}

Another remarkable point is that even during relatively long periods
of very low transfer rates (Fig.\,\ref{exam-3}, days\,4700--\,5000) the
disc does not stop its outburst activity but produces only short
outbursts with slowly decreasing amplitudes and increasing quiescence
intervals. This confirms the findings of King \& Cannizzo (1998).

Fig.\,7 shows 500 days of our simulation in which the adopted mass transfer
from the
secondary varies strongly on a timescale of roughly 20 days.
This is the most frequent case during our calculation but there are rare
periods of nearly constant mass transfer.
Fig.\,8 shows the light curve of the fictitious system
during 500 days in which the mass transfer rate nearly
exactly equals its averaged value (top). As a result the light curves
computed with the real mass transfer rate and the averaged value look equal.

In summary, one can say that the variations of the mass transfer rate
leads the disc to switch between three states in which only long
outbursts occur ($\log(\dot{M}_{\mathrm{tr}})\geq 16.3$), one long
outburst is followed by one or two short outbursts
($16.3>\log(\dot{M}_{\mathrm{tr}})> 15.7$), and only short outbursts
occur ($\log(\dot{M}_{\mathrm{tr}})\leq15.7$).

In order to understand the described behaviour of the disc we take into
account the viscous
timescale $t_{\mathrm{v}}\sim\,R^2/\nu$
which gives an estimate of the timescale for a disc annulus to move a radial
distance $R$.
For the quiescent state $t_{\mathrm{v,c}}$ and the outburst state
$t_{\mathrm{v,h}}$ and with $R=R_{\mathrm{out}}$ (where
the mass transferred from the secondary is added to the disc) we obtain for the
viscous timescale
\begin{eqnarray}
t_{\mathrm{v,c}}\sim\,\frac{R^2}{\nu}\,\sim\,2000\,\mathrm{d}, \\
t_{\mathrm{v,h}}\sim\,\frac{R^2}{\nu}\,\sim\,15\,\mathrm{d}.
\end{eqnarray}
The mass added to the disc during quiecence is stored in the disc
because it moves inward on the long timescale $t_{\mathrm{v,c}}$ whereas
during an outburst even mass from outer regions
can reach the white dwarf within a viscous timescale. This makes it
possible that the mass accreted
onto the white dwarf during a long outburst can be up to roughly one third of
the disc mass ($\sim\,\mathrm{day}\,4220$). Therefore the disc can relax
to equillibrium
with the mass transfer rate in only one outburst in the case of high transfer
rates (Fig.\,4).

The prompt response of the disc to the sharp decline in the mass transfer rate
(Fig.\,5) can be
understood in the same way:
due to the short viscous time ($t_{\mathrm{v,h}}$) the disc accretes a
substantial fraction of the disc mass ($\sim\,1/4 M_{\mathrm{disc}}$) during the last long outburst which immediately prevents
long outbursts when the mass transfer rate becomes low ($\sim\,\mathrm{day}\,4340$).

Finally, the long period of low transfer rates (Fig.\,6) do not prevent
outbursts because the mass accreted during the short outbursts is only a few
percent of the disc mass. Therefore the disc relaxes to the mass transfer
rate on a longer timescale.

\subsection{The mass transfer and mass accretion rates }

\begin{figure}
\includegraphics[angle=270,width=9cm]{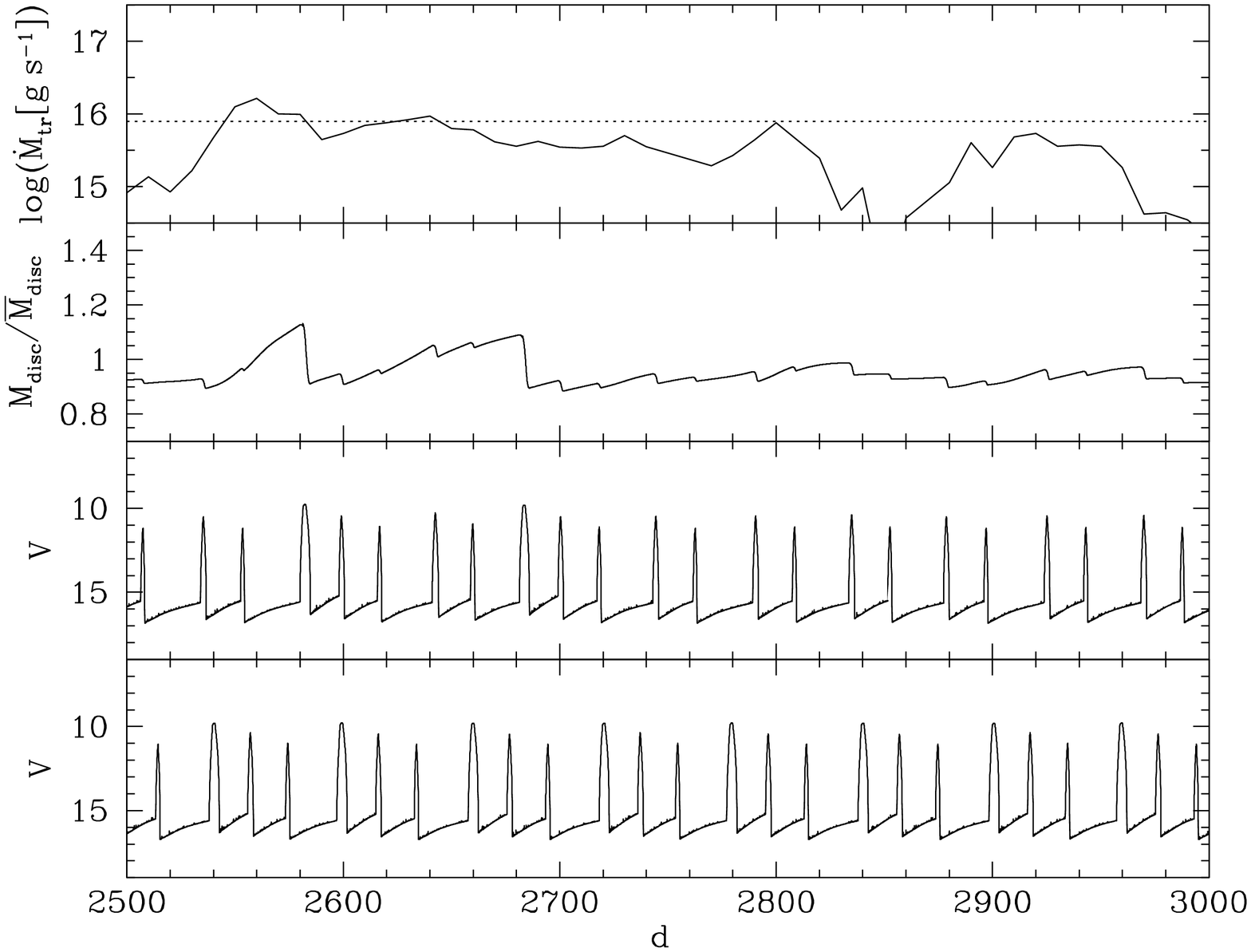}
\caption[]{\label{mtracc-1} The same as Fig.\,\ref{exam-1} but a
snapshot of the simulation where the transfer rate strongly varies. }

\includegraphics[angle=270,width=9cm]{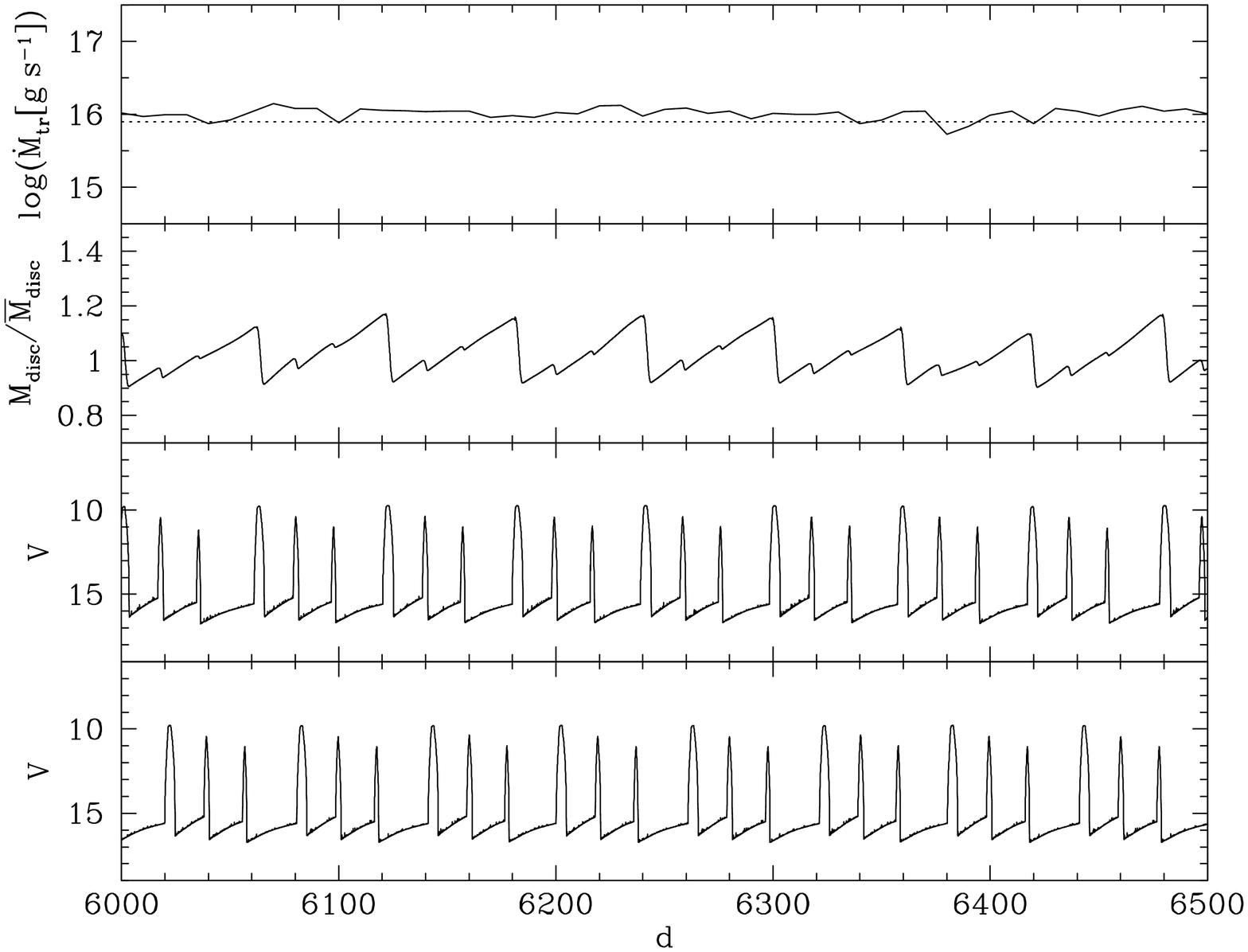}
\caption[]{\label{mtracc-2} The same as Fig.\,\ref{exam-1} but a
snapshot of the simulation where the transfer rate nearly stays
constant.}

\end{figure}
An important point in understanding the physics of accreting binaries
is to know how far the outburst behaviour and hence the
resulting light curves depend on real variations of the mass tranfer
rate.

To answer this question, we compare the averaged mass accretion rate
onto
the white dwarf with the mass transfer
rate. Figure\,9 shows that the time in which the disc relaxes to an
equilibrium with the mass transfer rate depends on the occurence of long
outbursts: when the accretion rate is averaged over 20\,days (dotted line
in Fig.\,9) the mass transfer and the accretion rate correspond roughly
only during periods where only long outbursts occur (days\,40\,--\,150,
see also Fig.\,4)
but for the periods where the disc goes through a cycle of
short and long outbursts the accretion rate has to be averaged over
 60 days to match the mass transfer rate (days\,150\,--\,500).
If the mass transfer rate drops steeply and stays in a low--state
(days\,4700\,--\,5000 in Fig.\,10, see also Fig.\,6), the accretion rate needs more than 60 days
to follow this behaviour, because the disc produces only short outbursts
in which only a small percentage of the disc mass is involved.

In order to make this plausible
we give a relaxation timescale $t_{\mathrm{r}}$ as the ratio of the viscous
timescale in outburst
with the relative mass fraction accreted during
an outburst:
\begin{equation}
t_{\mathrm{r}}\,\sim\,t_{\mathrm{v,h}}\frac{M_{\mathrm{disc}}}{\Delta\,{M}_{\mathrm{disc}}}
\end{equation}
For high mass transfer rates this
timescale is around $\,70\,\mathrm{days}$ and
so the correspondence of the averaged accretion rate and the
mass transfer
rate in Fig.\,9. is not surprising. In the case of low transfer rates
(Fig.\,10, day 4700--5000) this timescale is longer
($\sim\,300\,\mathrm{days}$) because only roughly 5 percent of the disc
mass are
accreted during an outburst.

\subsection{Dependence on the primary mass}
In the numerical experiment above, we have assumed an average white
dwarf mass for the primary in AM\,Her.  The literature holds a large
spectrum of white dwarf mass estimates for AM\,Her,
$M_\mathrm{wd}=0.39\,M_{\sun}$ (Young \& Schneider 1981),
$M_\mathrm{wd}=0.69\,M_{\sun}$ (Wu et al.1995),
$M_\mathrm{wd}=0.75\,M_{\sun}$ (Mukai \& Charles 1987),
$M_\mathrm{wd}=0.91\,M_{\sun}$ (Mouchet 1993) and
$M_\mathrm{wd}=1.22\,M_{\sun}$ (Cropper et al. 1998).  Based on the
observed ultraviolet spectrum of AM\,Her and on its well-established
distance, G\"ansicke et al. (1998) estimated
$R_\mathrm{wd}\approx1.1\times10^9$\,cm, and, using the
Hamada-Salpeter (1961) mass-radius relation for carbon cores,
$0.35\,M_{\sun}\,<\,M_\mathrm{wd}\,<\,0.53\,M_{\sun}$.  As the
Hamada-Salpeter mass-radius relation is valid for cold white dwarfs,
the finite temperature $\approx20\,000$\,K of the white dwarf in
AM\,Her would allow also somewhat higher masses,
$M_\mathrm{wd}\approx0.65M_{\sun}$, which is very close to the {\em
average} mass of field white dwarfs, $0.6M_{\sun}$, that we used.

Even though we exclude a massive white dwarf based on the
observational evidences, we repeated our simulation with
$M_\mathrm{wd}=1.0M_\mathrm{\sun}$ in order to test
the sensitivity of our results.

\begin{figure}
\includegraphics[angle=270,width=9cm]{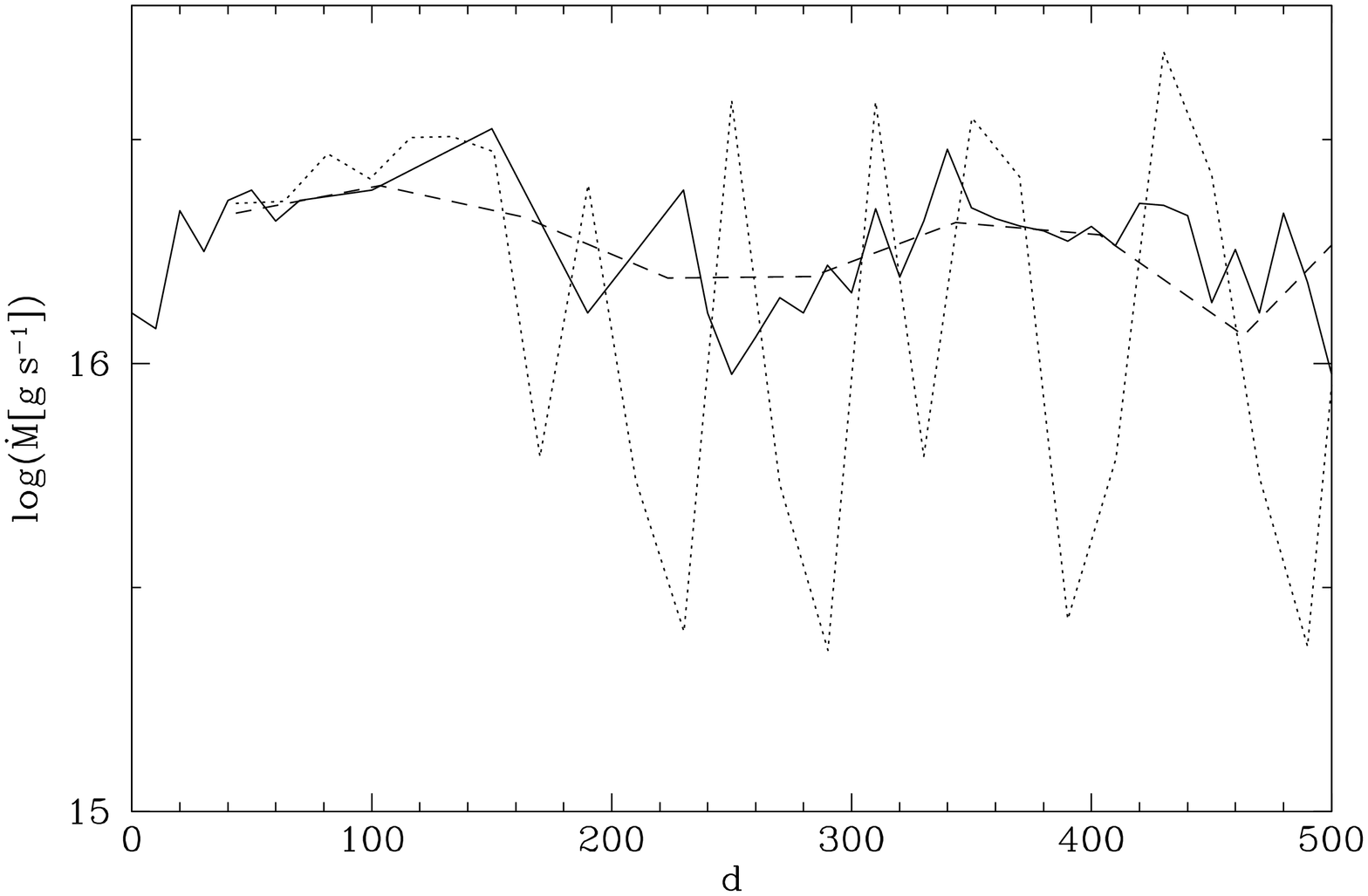}
\caption[]{\label{mtracc-2}
The mass transfer rate (solid line) and
the accretion rate onto the white dwarf as function of time. The
accretion rate is averaged over 60 days (dashed line) and averaged
over 20 days (dotted line). }

\includegraphics[angle=270,width=9cm]{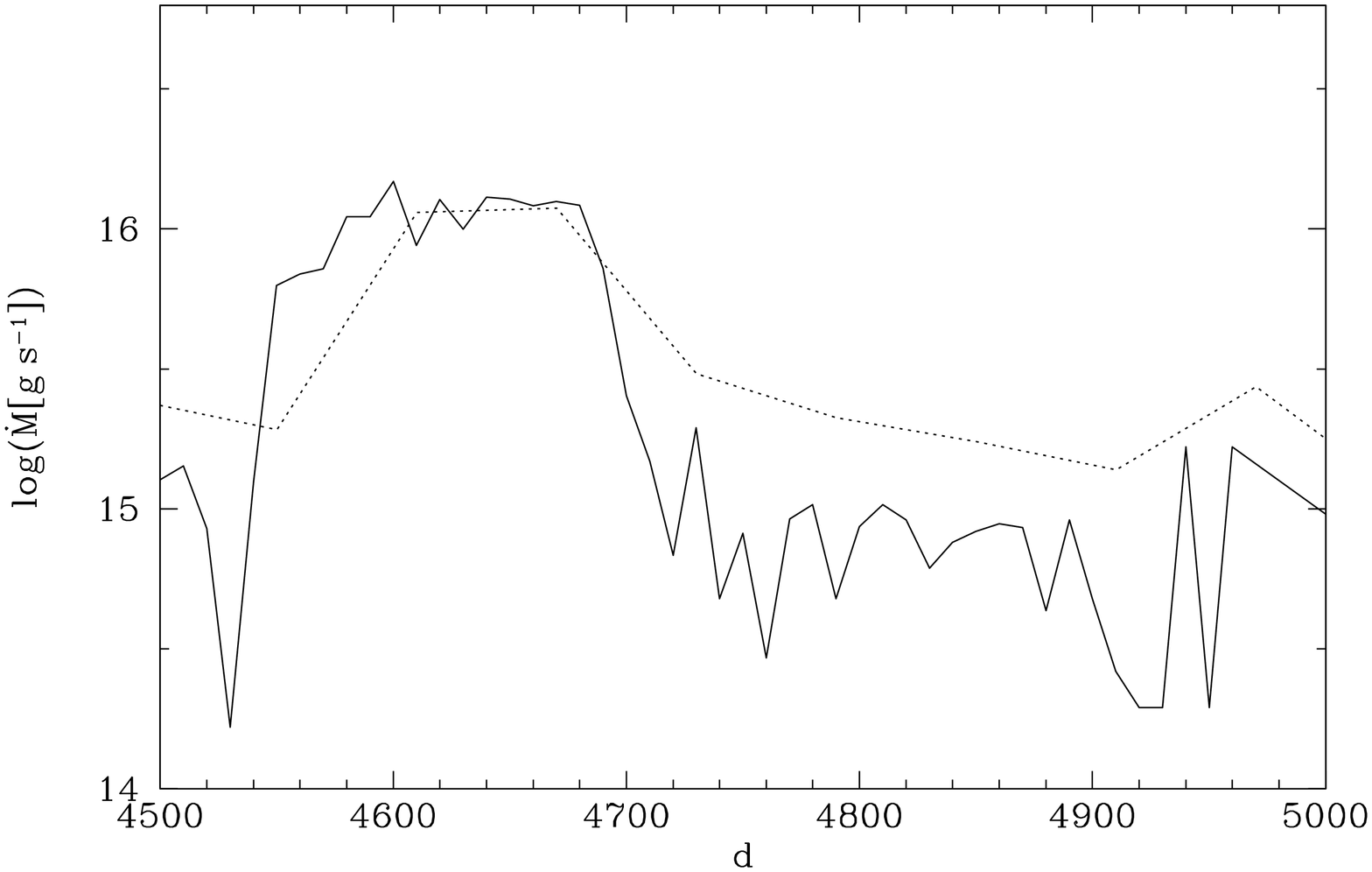}
\caption[]{\label{mtracc-2b}
The mass transfer rate (solid line) and
the accretion rate onto the white dwarf as function of time. The
accretion rate is averaged over 60 days (dashed line).}
\end{figure}

In a first step, we recompute $\dot M_\mathrm{tr}(t)$ from
Eq.\,(\ref{mtr}) with $M_{\mathrm{wd}}=1.0\,M_{\sun}$ and a corresponding
$R_\mathrm{wd}=5.4\times10^8$\,cm.  The resulting mean accretion rate
is $3.0\times10^{15}\mathrm{g\,s^{-1}}$, a factor 2.6 lower than
before. Then, we simulated once more 7000 days of disc evolution with
the new $\dot M_\mathrm{tr}(t),\, \alpha_{\mathrm{h}}=0.1,\, \alpha_{\mathrm{c}}=0.02$ and $R_{\mathrm{out}}=2.8\times10^{10}\,\mathrm{cm}$.

\begin{figure}
\includegraphics[angle=270,width=9cm]{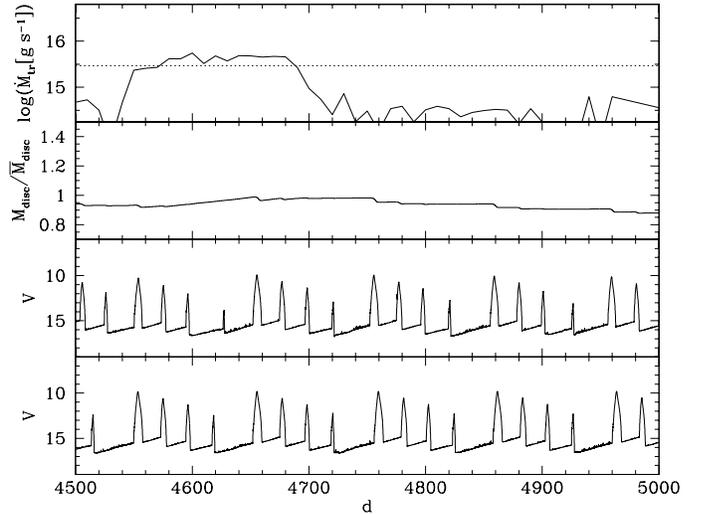}
\caption[]{\label{exam-4} The same as in Fig.\,\ref{exam-3} but
assuming a more massive primary
$M_{\mathrm{wd}}=1.0\,M_{\mathrm{\sun}}$ }
\end{figure}

In Fig.\,\ref{exam-4} we show 500 days of our calculation with
$M_\mathrm{wd}=1.0\,M_\mathrm{\sun}$. The disc produces only short
outbursts and the outburst cycle of four outbursts with decreasing
amplitude is hardly changed even by drastic variations of the mass
transfer rate (day\,4700).

The different responses that our fictitious dwarf novae with
$0.6M_{\sun}$ and $1.0M_{\sun}$ white dwarfs show to the variable mass
transfer rate are easy to understand: both the increased
primary mass and the decreased radius of the white
dwarf reduce - as mentioned above - the derived average mass transfer
rate. 

In addition, the outer disc radius of the fictitious system
with $M_\mathrm{wd}=1.0M_{\sun}$ increases.
The disc becomes more massive, i.e.
$\overline{M}_{\mathrm{disc}}=4.59\times10^{23}\mathrm{g}$, because of its
increased size. 

Due to the reduced mass transfer rate and the increased disc size, the
heating waves are able to reach the outer edge of the disc only during
the first outburst of the outburst cycle and only in cases where the mass
transfer rate
is high ($\log\,\dot{M}_{\mathrm{tr}}\geq\,16.1$). Even in this rare situation
the disc stays only a few days in the hot state and accretes only a
small fraction of the disc mass ($\sim 1/8\,M_{\mathrm{disc}}$).
For lower transfer rates (the extremly more frequent case shown in Fig.\,11)
the heating front gets reflected before it has
reached the outer edge of the disc. Hence, only a small percentage of the
disc mass is involved in any outburst.
Therefore, and due to the longer viscous timescale, the relaxation timescale
given in Eq.\,(10) is always larger than a few years.

Summing up, the adopted primary mass and the average accretion rate 
play an important role on the influence that the variable mass
transfer rate has on the outburst behaviour.

\section{A note on the disc limit-cycle model}
A number of new features have been added to the  well known
limit-cycle model (Meyer \& Meyer-Hofmeister 1981; Smak 1982; Cannizzo, Gosh \& Wheeler 1982).
Hameury et al. (1998, 1999, 2000) have shown how the light
curves change if the disc size is allowed to vary and which effects
irradiation has on the outburst behaviour. Additionally these authors
discussed other sources of uncertainties such as the tidal torque and the
evaporation of the inner parts of
the disc (Meyer \& Meyer-Hofmeister 1994). They concluded that all these effects must be
included in order to obtain meaningful physical information on
e.g. the viscosity from the comparison of predicted and observed
light curves.  Another point of importance is the
interaction of the accretion stream leaving the secondary star and the
accretion disc. Schreiber \& Hessman (1998) tested the
influence of stream overflow on the disc evolution in dwarf
novae. They found that significant stream overflow can lead to reversion 
of the inward-travelling cooling front and create an outward-travelling heating front. This behaviour would produce small dips in the light curve
during the declining phase.

The model we used here contains neither irradiation nor allows
it the disc radi to vary. The stream mass is simply deposited in a
small Gaussian distribution near the outer edge of the disc.
Therefore we do not attempt at present a comparison with observed
light curves.
Nevertheless, all calculated light curves with constant
mass transfer from the secondary relax in a quasi-stationary outburst
cycle (of one or more outbursts) which repeat periodically.
Our model shows
in a qualitative way that real mass transfer variations may have a
dominant influence on the outburst behaviour at least in systems
with relatively small discs and strong mass transfer.  

\section{Discussion and conclusions}
There is no reason to assume that the mass transfer variations of the
secondary observed in
AM\,Her are not present in non-magnetic systems.  Our numerical
experiment including realistic variations of the mass transfer rate in
a dwarf nova system is, therefore,  a significant step towards a better
understanding of dwarf nova light curves, and, thereby, of the
underlying disc limit cycle.

The light curve produced by our fictitious system switches between
three states depending on the actual mass transfer rate. High
transfer rates lead to only long outbursts where the entire disc is
transformed into the hot state. If the transfer rate is near the
average value the disc goes through a cycle of three outbursts, 
one long outburst followed by two short ones. Even long periods
of low transfer rates do not force the disc to stop its outburst
activity: long outbursts are suppressed and the duration of
quiescence increases but the disc always produces short
outbursts. From this follows that the low-states of VY Sculptoris
stars (a subgroup of novalike variables) could not be caused by low transfer
rates alone (see also Leach et al. 1999).

We find that in our fictitious system the mass accreted during an
outburst cycle is dominated by the course of the mass transfer
rate if the mass transfer rate varies significantly. The disc always
relaxes to equilibrium with the mass input from the secondary. Thus, our
experiment strongly supports King \& Cannizzo's (1998) claim that dwarf
nova accretion discs are probably never in a stationary state but are
constantly adjusting to the prevailing value of
$\dot{M}_{\mathrm{tr}}$. Only during periods where the mass transfer
rate is nearly exactly constant the disc periodically repeat the quasy
stationary outburst cycle. Such periods are rare but occur in AM\,Her.

The strong influence of the mass transfer rate on the
outburst behaviour of the fictitious system clearly indicates that
probably most (if not all) the deviations from periodic outburst cycles seen
in the
light curves of dwarf novae are caused by variations of the
mass transfer rate.

\begin{acknowledgement}
We thank Daisaku Nagami and the referee
for their helpful comments and suggestions. 
MRS would like to thank the Deutsche Forschungsgemeinschaft for
financial support (Ma\,1545\,2-1). BTG thanks for support from the DLR
under grant 50\,OR\,99\,03\,6.
\end{acknowledgement}


\end{document}